\documentclass[a4paper]{jpconf}
\usepackage{graphicx}
\begin{document}
\title{Air Shower Measurements in
Karlsruhe\footnote{Contribution to the 80th anniversary of 
late Georgii Borisovich Khristiansen, the discoverer of the knee 
in the cosmic ray energy spectrum. He supported the start of
air shower investigations in Karlsruhe with keen interest 
and was a great help for us at that time.}}

\author{Andreas Haungs}

\address{Institut\ f\"ur Kernphysik, Forschungszentrum Karlsruhe,
76021~Karlsruhe, Germany}

\ead{haungs@ik.fzk.de}

\begin{abstract}
The Karlsruhe multi-detector set-ups KASCADE, KASCADE-Grande, and 
LOPES aim on measurements of cosmic rays in the energy range of the 
so called knee between $10^{14}\,$eV and $10^{18}\,$eV.
The multidimensional analysis of the air shower 
data measured by KASCADE indicates a distinct knee in the energy 
spectra of light primary cosmic rays and an increasing dominance 
of heavy ones towards higher energies. 
This provides, together with the results of large scale 
anisotropy studies, implications for discriminating astrophysical 
models of the origin of the knee. To improve the reconstruction 
quality and statistics at higher energies, where the knee of the 
heavy primaries is expected at around 100 PeV, KASCADE has been 
extended by a factor 10 in area to the new experiment KASCADE-Grande. 
LOPES is located on site of the KASCADE-Grande experiment. It  
measures radio pulses from extensive air showers with the goal 
to establish this renewed detection technique for future large 
scale experiments. 
\end{abstract}

\section{Introduction}

\begin{figure*}[t]
\centering
\vspace*{0.1cm}
\includegraphics[width=160mm]{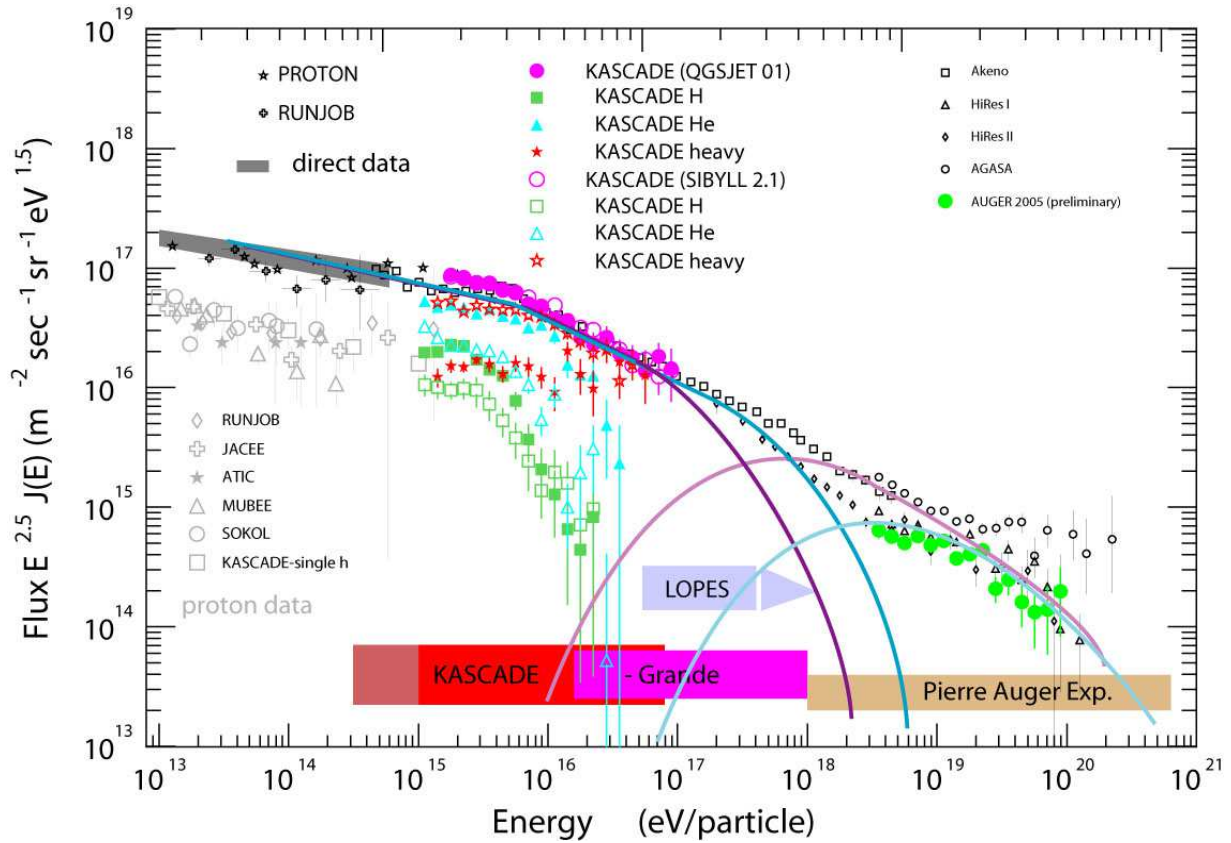}
\caption{\label{spectrum}Primary cosmic ray flux. 
Results of some experiments (in particular KASCADE) 
for the all-particle spectrum as well as 
for spectra of individual mass groups, in particular for protons 
are displayed. 
Full lines show possible scenarios for the origin of a second knee 
in the spectrum (explanation see text chapter~4).}
\label{knee}
\end{figure*}

\vspace*{0.5cm}
The all-particle energy spectrum of cosmic rays exhibits 
a distinctive discontinuity at few PeV, known as the knee, 
where the spectral index
changes from $-2.7$ to approximately $-3.1$~(Fig.~\ref{knee}). 
This feature has been discovered half a century ago by
German Kulikov and George Khristiansen of the Moscow State
University~\cite{knee}
within studies of the intensity spectrum of  
the content of charged particles of extensive air showers, 
which roughly reflects the primary 
energy. 
At that energy direct measurements are hardly possible due to 
the low flux. 
Thus indirect measurements observing extensive air showers (EAS) 
attempt to reveal the structure of the spectrum.  \\

The key questions of the origin of this knee are still 
not convincingly solved.
Astrophysical scenarios like the change of the acceleration 
mechanisms at the cosmic ray sources 
(supernova remnants, pulsars, etc.) 
or effects of the transport mechanisms inside the Galaxy (diffusion 
with escape probabilities) are conceivable for the origin of the knee
as well as particle physics reasons like a new kind of hadronic 
interaction inside the atmosphere or during the transport through 
the interstellar medium. An overview on the current zoo of these 
theoretical models were recently given in~\cite{joerg}.
It is obvious that only detailed measurements covering the full 
energy range of the knee from $10^{14}\,$eV to $10^{18}\,$eV
and analyses of the primary energy spectra for the different incoming 
particle types can validate or disprove some of 
these models. 
Despite EAS measurements with various different 
experimental set-ups in the 
last decades this demand could never accomplished, mainly due to 
the weak mass resolution of the measured shower 
observables~\cite{rpp}. \\

The multi-detector system KASCADE-Grande (KArlsruhe Shower Core and Array 
DEtector and Grande array)~\cite{Navar04}, 
approaches this challenge by 
measuring as much as possible independent information from
each single air-shower event. 
In 2003, the original KASCADE experiment~\cite{kas}, which is
optimized for the energy range of $10^{14}\,$eV to $10^{16.8}\,$eV 
has been extended in 
area by a factor 10 to the new
experiment KASCADE-Grande.
KASCADE-Grande allows now a full coverage of the energy range 
around the knee, including the possible second 
knee at energies just below $10^{18}\,$eV. 

With its capabilities KASCADE-Grande is also the ideal testbed for 
the development and calibration of new air-shower
detection techniques like the measurement of EAS radio
emission, which is performed in the frame of the LOPES 
project~\cite{lopes}.
LOPES is designed as digital radio interferometer using high bandwidths 
and fast data processing and profits from the reconstructed air shower 
observables of KASCADE-Grande. 

\section{The Multi-Tool Box: 
KASCADE, KASCADE-Grande, and LOPES}

\begin{figure}[hb]
\centering
\begin{minipage}{75mm}
\centering
\includegraphics[width=72mm]{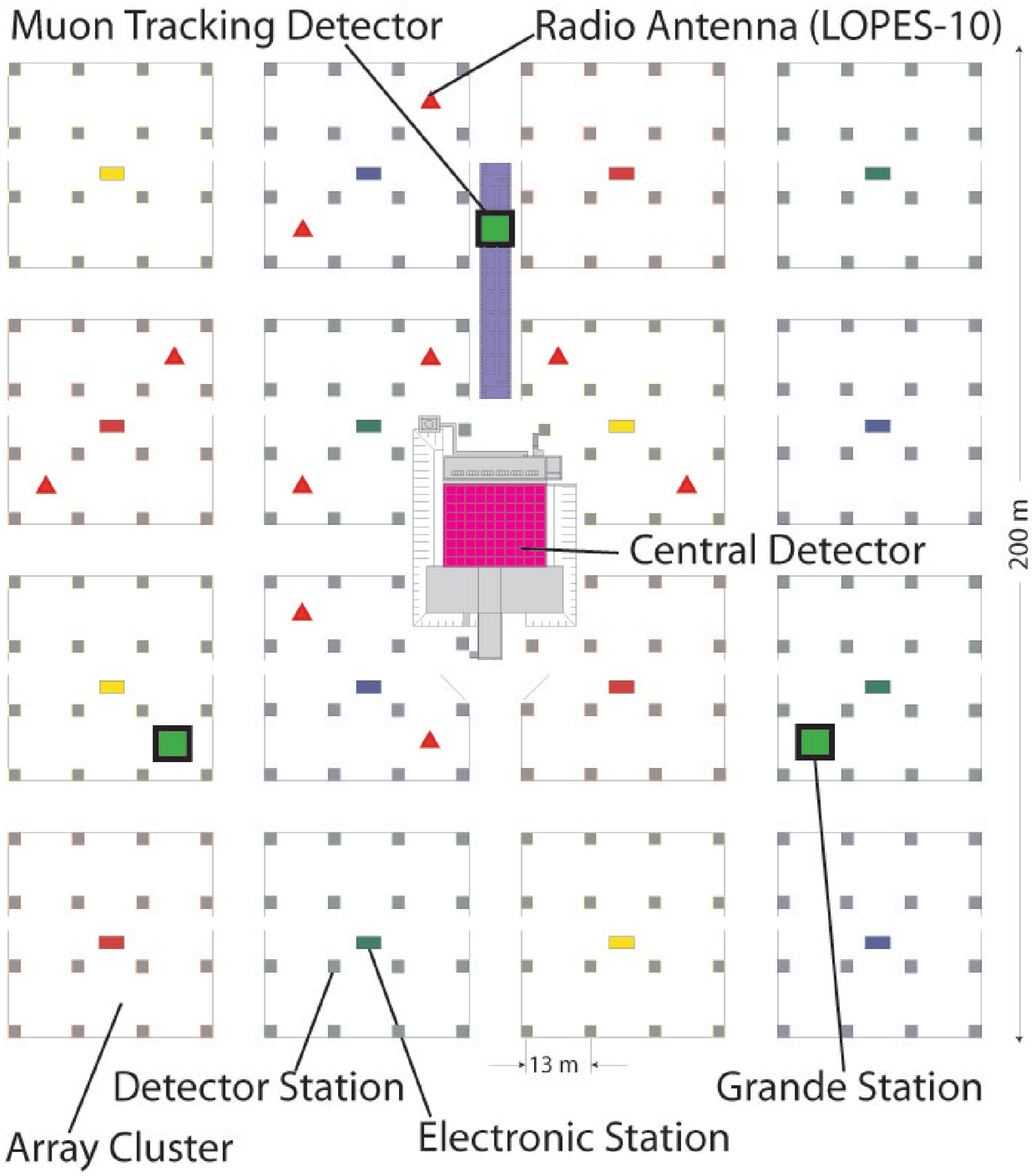}
\caption{\label{KASCADE_radio}
Sketch of the KASCADE experiment: 
field array, muon tracking and 
central detector. 
The outer 12 clusters of the array consists of $\mu$- and
$e/\gamma$-detectors, the inner 4 clusters of $e/\gamma$-detectors,
only. The locations of 10 radio antennas of LOPES~10 
are also displayed, as well as three 
stations of the Grande array.}
\end{minipage}\hspace{10mm}%
\begin{minipage}{75mm}
\centering
\includegraphics[width=75mm]{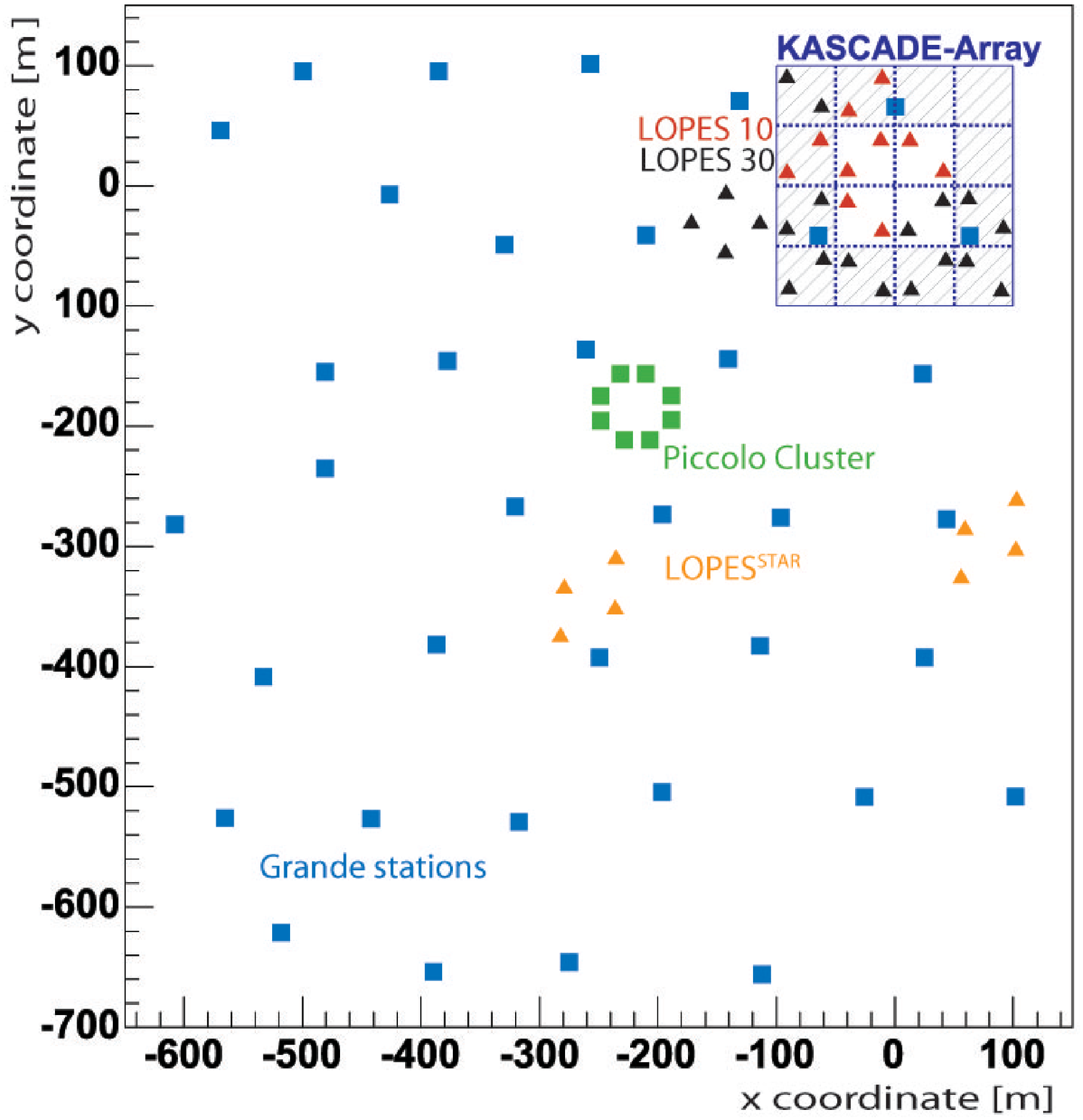}
\caption{Sketch of the KASCADE-Grande -- LOPES 
experiment: The KASCADE array, the distribution of the 
37 stations of the Grande
array, and the small Piccolo cluster for fast trigger purposes are
shown. The location of the 30 LOPES
radio antennas is also displayed as well as the LOPES$^{\rm STAR}$ 
antennas. \label{grande}}
\end{minipage} 
\end{figure}

\begin{figure}[ht]
\centering
\begin{minipage}{85mm}
\centering
\includegraphics[width=85mm]{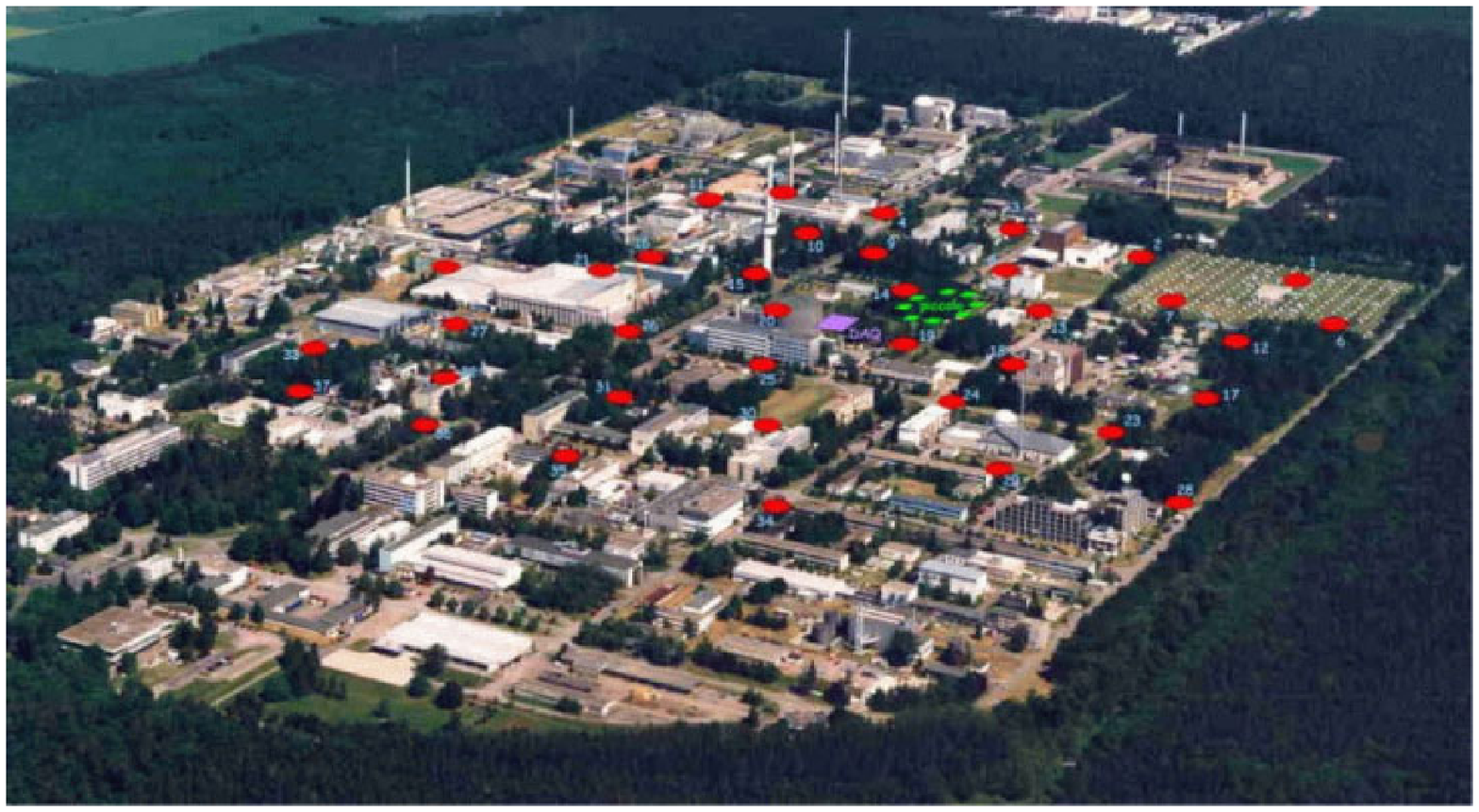}
\caption{\label{fzk}
Photograph of the Forschungszentrum Karlsruhe. 
KASCADE is situated in the Northeastern corner of the center; 
the Grande stations are indicated by red dots, the Piccolo array by
green dots.}
\end{minipage}\hspace{10mm}%
\begin{minipage}{65mm}
\centering
\includegraphics[width=65mm]{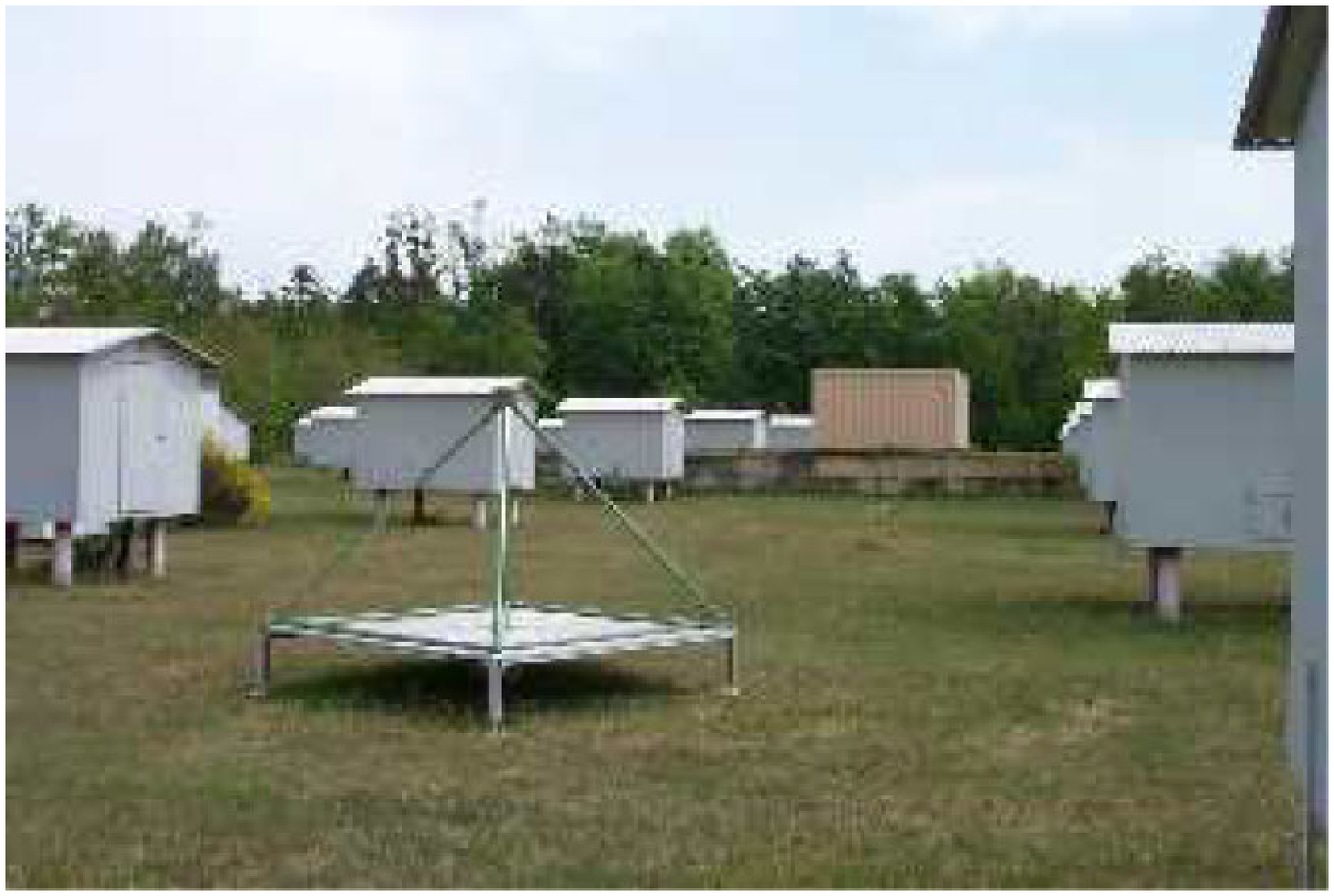}
\caption{Photograph of the experimental set-up in Karlsruhe with
several KASCADE array stations. 
In the foreground a LOPES antenna is seen, in the background a
Grande station on top of the underground muon tracking detector.
\label{lopes}}
\end{minipage} 
\end{figure}

\vspace*{0.5cm}
These experiments 
(Figs.~\ref{KASCADE_radio},~\ref{grande},~\ref{fzk},~\ref{lopes}) 
are located at the 
Forschungszentrum Karlsruhe, Germany, 
(49.1$^\circ$n, 8.4$^\circ$e, 110$\,$m$\,$a.s.l.) and measure 
extensive air showers in a primary energy range 
from $100\,$TeV to $1\,$EeV.
Their combination provides multi-parameter measurements on a 
large number of observables of the shower components:
Electrons, muons at 4 energy thresholds, hadrons, and the
radio emission.
The main detector components are the KASCADE array, 
the Grande array, and the LOPES antenna array. \\
\vspace*{0.1cm}

The KASCADE array measures the 
total electron and muon numbers ($E_{\mu,{\rm kin}}>230\,$MeV)
of the shower separately using an array of 252 detector stations 
\vspace*{0.1cm}
containing shielded and unshielded detectors at the same place 
in a grid of $200 \times 200\,$m$^2$ (Fig.~\ref{KASCADE_radio}).
The excellent time resolution of these detectors allows also decent 
investigations of the arrival directions of the showers in searching
large scale anisotropies and, if existent, cosmic ray point sources. 
The KASCADE array is optimized to measure EAS in the energy range of 
100 TeV to 80 PeV.

A muon tracking detector measures the incidence 
angles of muons relative to the shower arrival 
direction. These measurements provide a sensitivity to the
longitudinal development of the showers. 
The hadronic core of the shower is measured by a $300\,$m$^2$ 
iron sampling calorimeter installed at the KASCADE central detector. 
And three further components offer additional 
valuable information on the penetrating muonic component 
at different energy thresholds. 
The complementary information of the showers measured by the 
central and the muon tracking detectors 
is predominantly being used for a better understanding of the 
features of an air-shower and for tests and improvements 
of the hadronic interaction models underlying the 
analyses~\cite{isv04}.  \\

The multi-detector concept of the KASCADE experiment, which is 
operating since 1996 has been translated to higher primary 
energies through KASCADE-Grande~\cite{KG-cris}.
The 37 stations of the Grande Array (Fig.~\ref{grande})  
extend the cosmic ray measurements up to primary 
energies of \mbox{1 EeV}. 
The Grande stations, \mbox{10 m$^2$} of plastic scintillator 
detectors each, are spaced at approximative by \mbox{130 m} 
covering a total area of \mbox{$\sim$ 0.5 km$^2$}. In addition,
a small cluster of stations (Piccolo) 
close to the center of the Grande array
is installed in order to provide a fast trigger to the muon 
detection systems of the original KASCADE array. \\

For the calibration of the radio signal emitted by the air shower in
the atmosphere an array of first 10 and meanwhile 30 dipole antennas 
(LOPES) is set up on site of the 
KASCADE-Grande experiment~\cite{spie,lopes-texas}. 
The basic idea of the LOPES (= LOFAR prototype station) project 
is to build an array of relatively simple, quasi-omnidirectional 
dipole antennas, where the received waves are digitized and sent 
to a central computer. This combines the advantages of low-gain 
antennas, such as the large field of view, with those of 
high-gain antennas, like the high sensitivity and good background 
suppression. 
With LOPES it is possible to store the received data stream for a 
certain period of time, i.e.~at a detection of a transient 
phenomenom like an air shower retrospectively a beam in the desired 
direction can be formed.
The air shower experiment KASCADE-Grande provides a trigger of 
high-energy events and additionally with its direction 
reconstruction a starting point for the radio data analyses and 
the beam forming. 
In the current status LOPES 
operates 30 short dipole radio 
antennas (LOPES-30) having now an absolute calibration.
Data of the first 10 antennas 
forming LOPES-10 have so far been analyzed.
All LOPES-30 antennas are deployed in east-west direction, 
measuring the east-west polarization, only.
In a new measuring campaign, also dual-polarized LOPES antennas 
will be used.
In addition, LOPES runs a field of logarithmic-dipole-antennas (LPDA) which are 
optimized for an application at the Pierre-Auger-Observatory, and for 
developing a self-trigger system (LOPES$^{\rm STAR}$)~\cite{gemmeke}. 
The layout is depicted in Fig.~\ref{grande}.
All the antennas operate in the frequency range of 
$40-80\,$MHz.
The read out window for each LOPES-30 antenna is $0.8\,$ms wide,  
the sampling rate is $80\,$MHz.
The geometry of the antenna and the aluminum ground screen give the 
highest sensitivity to the zenith and half sensitivity to zenith 
angles of $43^\circ - 65^\circ$, dependent on the azimuth angle.  
LOPES-30 data are read out if KASCADE-Grande triggers by a high 
multiplicity of fired stations, 
corresponding to primary energies above $\approx 10^{16}\,$eV. 
Such showers are detected at a rate of $\approx 2-3$ per minute.

\section{KASCADE: Light primaries drop a curtsy}

\vspace*{0.6cm}
{\centering
\begin{minipage}{120mm}
`An analysis of these and other data available in the literature
indicates that there is, very probably, an
irregularity in the shower
size distribution curve in the region between $10^6$ and $10^7$
particles.' \\
{\it from: On the size spectrum of extensive air showers (1959), G.V.
Kulikov and G.B. Khristiansen~\cite{knee}.}
\end{minipage}}
\vspace*{0.6cm}

\begin{figure}[hb]
\centering
\vspace*{0.1cm}
\includegraphics[width=110mm]{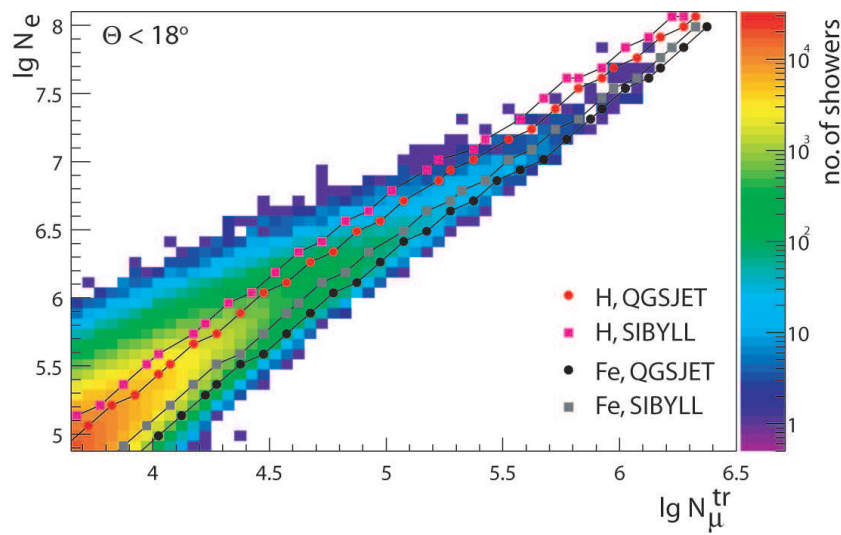}
\caption{Two dimensional electron ($N_e$)~vs.~muon 
($N_\mu^{\rm tr}\,=\,$number of muons within 40-200m core distance) 
number spectrum measured by the KASCADE array. 
The lines display the most probable values for proton
and iron primaries obtained by CORSIKA simulations employing 
different hadronic interaction models~\cite{isv04}.} 
\label{data}
\end{figure}

\begin{figure}[ht]
\centering
\begin{minipage}{75mm}
\centering
\includegraphics[width=7.5cm]{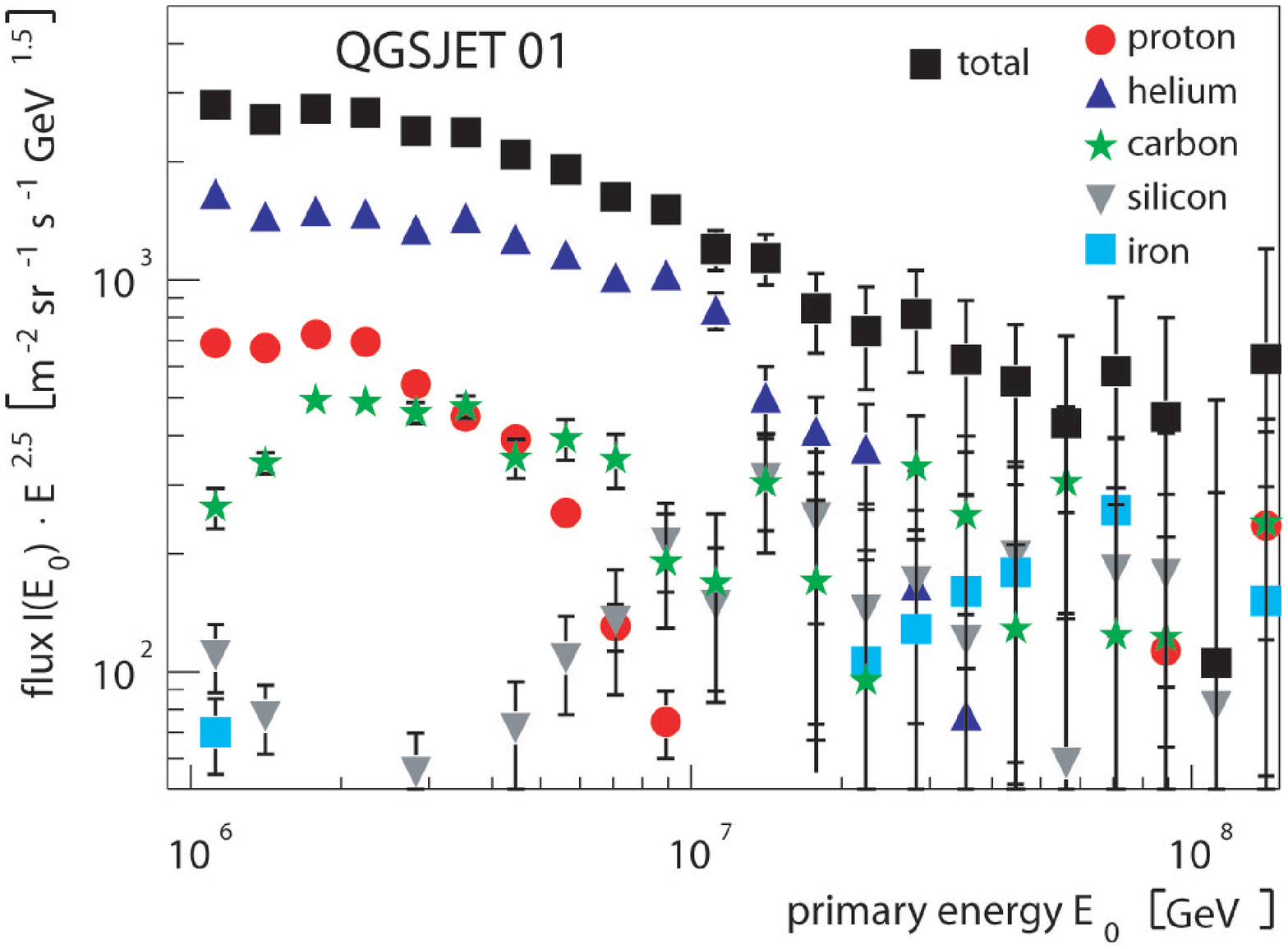}
\caption{Result of the unfolding procedure based
on QGSJET$\,01$~\cite{Ulric05}.}
\label{spectra-qgs}
\end{minipage}\hspace{10mm}%
\begin{minipage}{75mm}
\centering
\includegraphics[width=7.5cm]{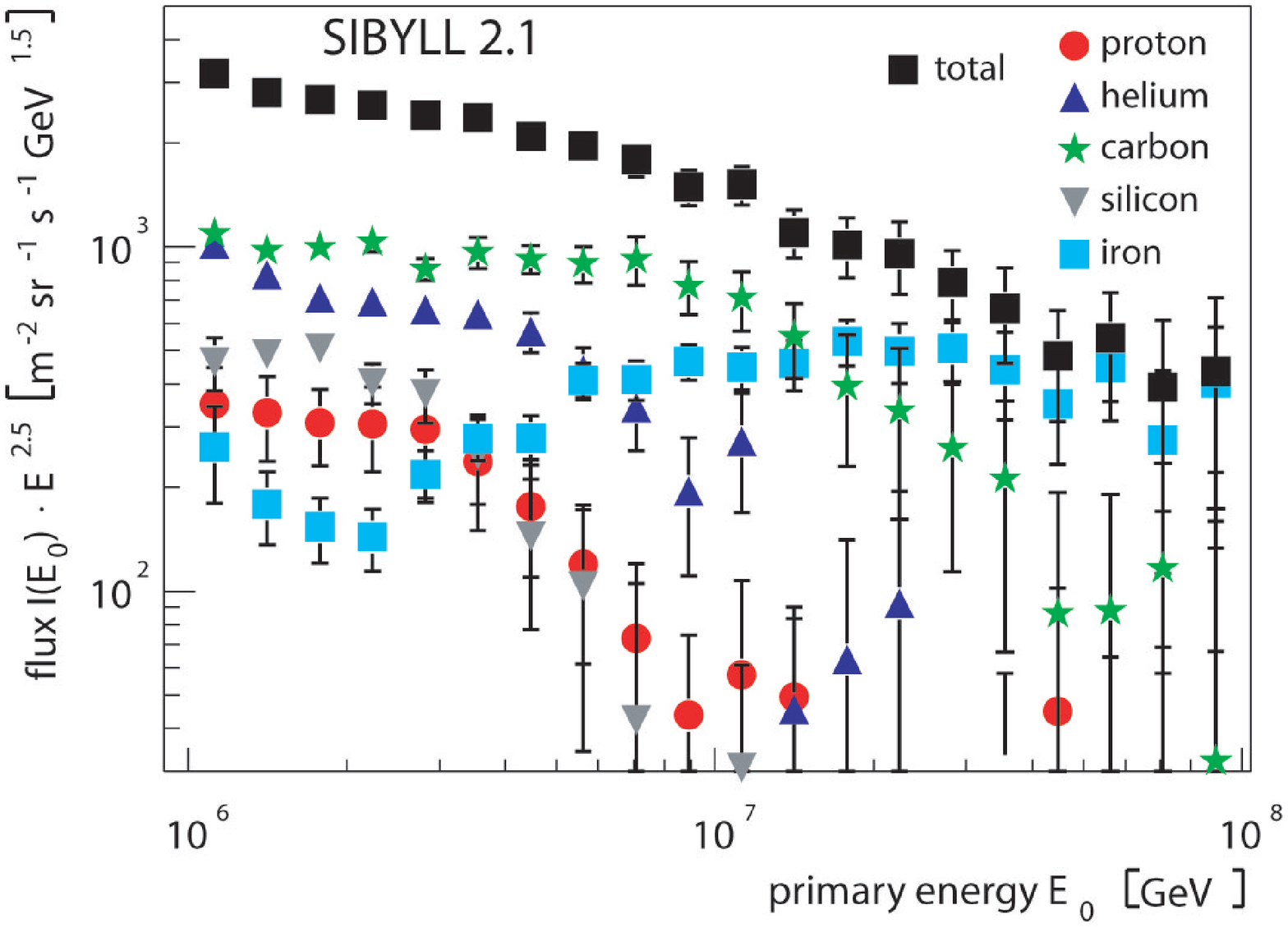}
\caption{Result of the unfolding procedure based
on SIBYLL$\,2.1$~\cite{Ulric05}.}
\label{spectra-sibyll}
\end{minipage} 
\end{figure}

The KASCADE data analyses aims to reconstruct the energy spectra 
of individual mass groups taking into account not only different
shower observables, but also their correlation on an event-by-event
basis.
The content of each cell of the two-dimensional spectrum of 
reconstructed electron number~vs.~muon number  
is the sum of contributions from the individual primary elements.
Hence the inverse problem 
{\small $g(y) = \int{K(y,x)p(x)dx}$} 
with {\small $y=(N_e,N_\mu^{\rm tr})$} (see Fig.~\ref{data}) 
and {\small $x=(E,A)$} 
has to be solved.
This problem results
in a system of coupled Fredholm integral equations of the form 
 {\small  $\frac{dJ}{d\,\lg N_e\,\,d\,\lg N_\mu^{tr}} = 
 \sum_A \int\limits_{-\infty}^{+\infty} \frac{d\,J_A}{d\,\lg E} 
  \cdot 
  p_A(\lg N_e\, , \,\lg N_\mu^{tr}\, \mid \, \lg E)
  \cdot 
  d\, \lg E $ }
where the probability $p_A$ is obtained by 
Monte Carlo simulations on basis of 
two different hadronic interaction models (QGSJET$\,01$~\cite{qgs}, 
SIBYLL$\,2.1$~\cite{sib}) as options embedded in 
CORSIKA~\cite{cors}. 
By applying these procedures (with the assumption of
five primary mass groups) to the experimental 
data energy spectra are obtained as displayed 
in Figs.~\ref{spectra-qgs} and~\ref{spectra-sibyll}. 
The results are also shown in Fig.~\ref{knee}, where 
the resulting spectra for primary oxygen, silicon,
and iron are summed up for a better visibility. \\

\begin{figure}[ht]
\centering
\begin{minipage}{60mm}
\centering
\includegraphics[width=60mm]{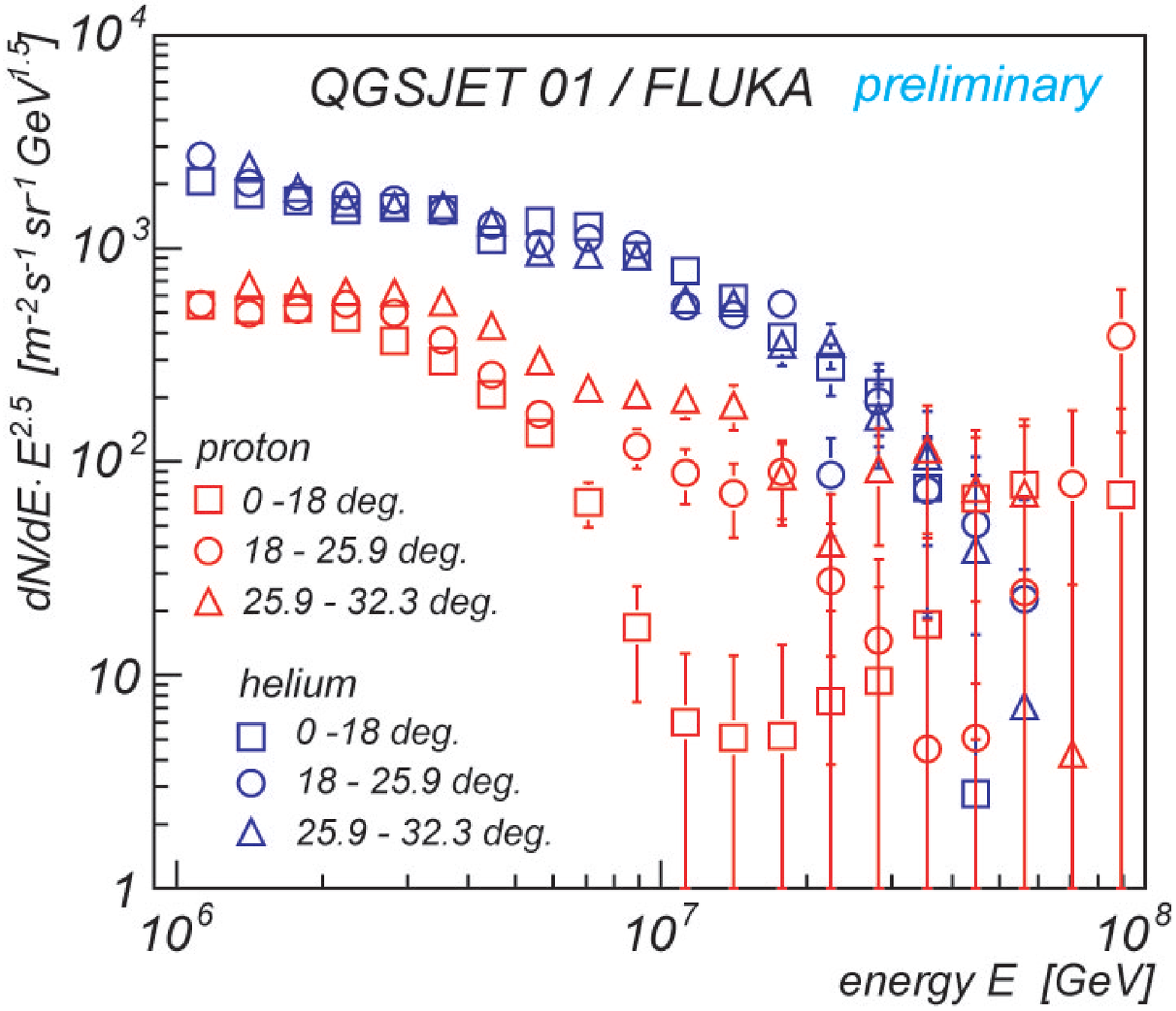}
\caption{\label{angles} Comparisons of the energy spectra of protons and Helium for different zenith angular ranges~\cite{Ulric06}.}
\end{minipage}
\hspace{10mm}
\begin{minipage}{85mm}
\centering
\includegraphics[width=85mm]{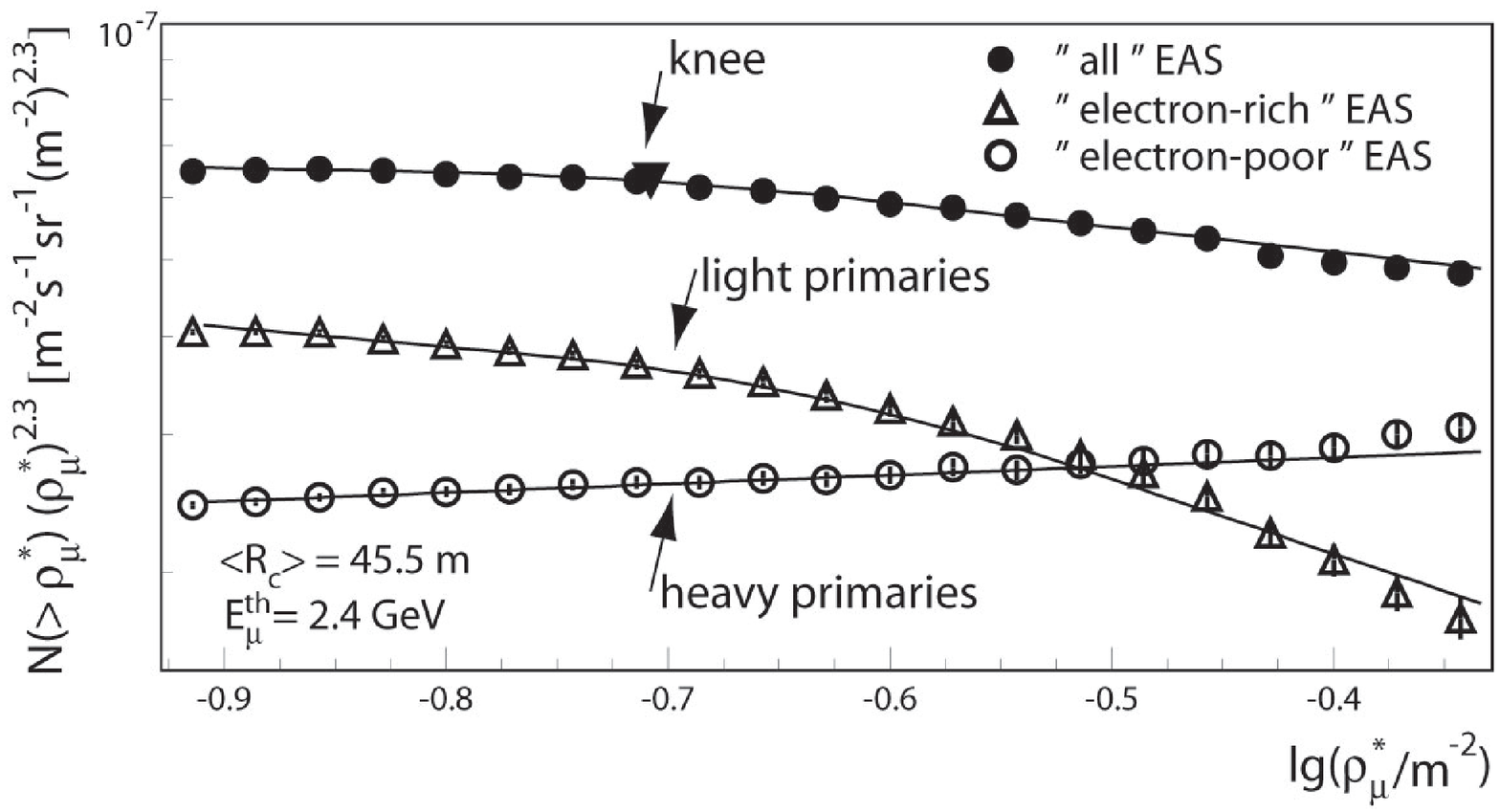}
\caption{\label{muon}Muon density spectra for different samples of EAS
as obtained by KASCADE measurements~\cite{muon}.}
\end{minipage}%
\end{figure}

A knee like feature is clearly visible in the all particle spectrum,
which is the sum of the unfolded single mass group spectra, 
as well as in the spectra of primary proton and helium.
\vspace*{0.1cm}
This demonstrates that the elemental composition of cosmic
rays is dominated by the light components below the knee and 
by a heavy component above the knee feature. Thus, the knee feature 
originates from a decreasing flux of the light primary 
particles~\cite{Ulric05}. Recently, the described analysis 
was repeated by using a different low energy interaction model 
(FLUKA~\cite{fluka}) and by comparing the resulting spectra of 
cosmic rays registered in different ranges of the zenith angle.
Fig.~\ref{angles} displays the results, where it is seen 
that the above mentioned findings are confirmed~\cite{Ulric06}. \\

Comparing the unfolding results based on the two different high-energy
hadronic interaction models QGSJet and SIBYLL, 
the model dependence when interpreting the data is obvious. 
Modeling the hadronic 
interactions underlies assumptions from particle physics theory and 
extrapolations resulting in large uncertainties, which are reflected 
by the discrepancies of the results presented here. 
In Fig.~\ref{data} the predictions of the $N_e$ and $N_\mu^{tr}$ 
correlation for the 
two models are overlayed to the measured distribution
in case of proton and iron primaries. 
It is remarkable that all four lines have a more or less parallel 
slope which is different from the data distribution. There, the 
knee is visible as kink to a flatter $N_e$-$N_\mu^{tr}$ 
dependence above $\lg N_\mu^{tr} \approx 4.2\,$. 
Comparing the residuals of the unfolded two dimensional 
distributions for 
the different models with the initial data set 
one can conclude
that at lower energies the SIBYLL model and at higher energies 
the QGSJET model are able to describe 
the correlation consistently, but none of the present models
gives a contenting description of the whole data set~\cite{isv04}. 

Crucial parameters in the modeling of hadronic 
interaction models which can be responsible for these inconsistencies 
are the total nucleus-air cross-section and the parts of the inelastic and 
diffractive cross sections leading to shifts of the position of the 
shower maximum in the atmosphere, and therefore to a change of the 
muon and electron numbers as well as to their correlation on single 
air shower basis. 
The multiplicity of the pion generation at all energies at 
the hadronic interactions during the air shower development is also
a `semi-free' parameter in the air-shower modeling as accelerator 
data have still large uncertainties.

Arbitrary changes of free parameters 
in the interaction models will change the 
correlation of all shower parameters. Tests using KASCADE observables,
which are measured independently of such used in the unfolding procedure,
may give further constraints, in particular by investigating correlations of 
the hadronic shower component with electron or muon numbers. 
The aim is to provide 
hints for the model builder groups how the parameters (and the theory) 
should be modified in order to describe all the data consistently.
\vspace*{0.1cm}
\vspace*{0.1cm}
\begin{figure}[ht]
\centering
\begin{minipage}{85mm}
\centering
\includegraphics[width=85mm]{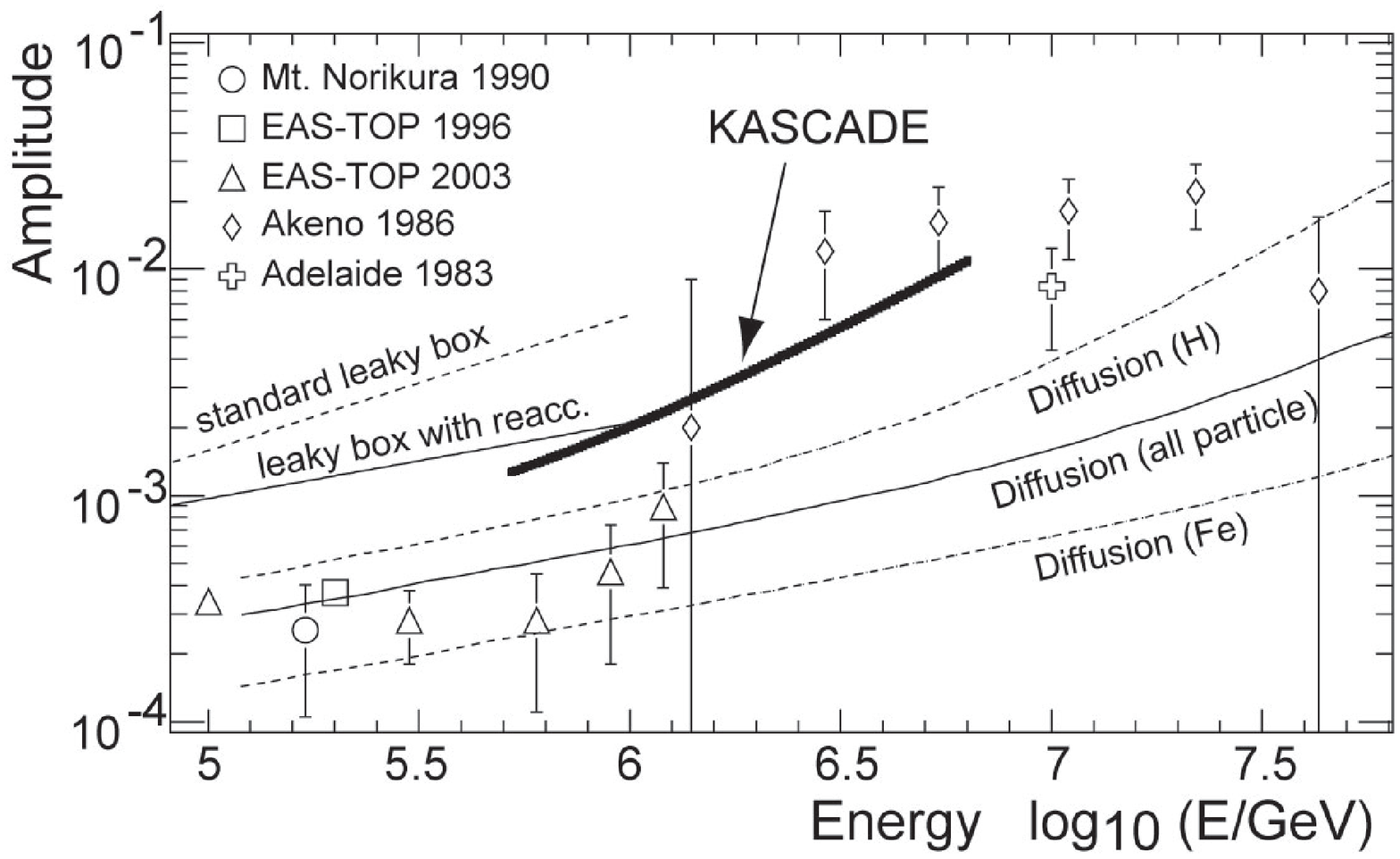}
\caption{\label{lsaniso} Rayleigh amplitude of the harmonic analyses 
of the KASCADE data~\cite{Gmaier1} (limit on a 95\% 
confidence level) compared to theory predictions~\cite{candia}.}
\end{minipage}
\hspace{10mm}
\begin{minipage}{60mm}
\centering
\includegraphics[width=60mm]{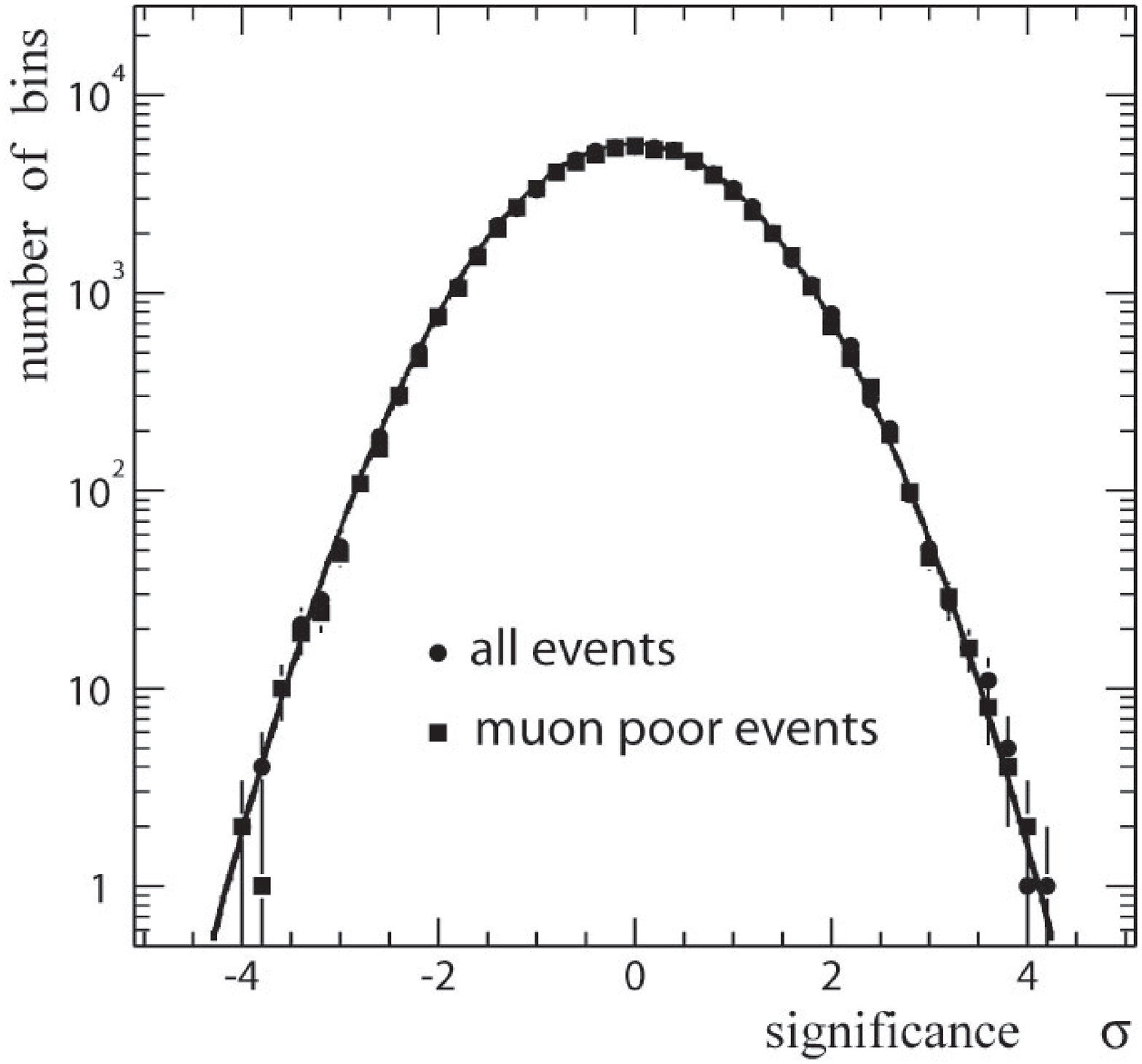}
\caption{\label{psaniso}Significance distributions for searching
point sources on the sky map seen by the KASCADE 
experiment~\cite{Gmaier2}.}
\end{minipage}%
\end{figure}

The applied method here is to evaluate the measured data relative to
simulations (including the detector response) 
of proton and iron primaries. The measured data points should fall 
between these extreme values, otherwise the simulations are unable
to describe this specific observable correlation. More 
direct comparisons
between data and simulations are not possible due to the unknown
composition of the primary particles generating the air showers.
This kind of tests is performed for a large set of interaction models 
available in the simulation package CORSIKA~\cite{cors}. 
Results~\cite{JphysG-hc,isv04} of such detailed investigations 
confirm the deficiencies of the hadronic interaction models. \\

In summary, 
there are still deficiencies of the hadronic interaction models 
(in particular in describing the high energy interactions) which 
are revealed by the high accuracy data of KASCADE. 
But combining all the correlations, the final observables of the 
showers have not to be different by more than $\approx 15$\% to 
be consistent with the data. This requires rather a fine-tuning 
of the free parameters in the simulation of the hadronic 
interactions than a need on new physics. \\

Neglecting somehow the uncertainties by the interaction models, 
KASCADE used an additional independent and more direct approach to
interprete the measurements (Fig.~\ref{muon}). 
By using three observables (one observable as energy identifier - 
the local muon density of high energy muons; 
and two observables as mass identifier - the ratio of electron to 
low energy muon number for dividing the whole EAS sample in a sample 
generated by light primaries and heavy primaries) 
KASCADE could impressively and in a nearly 
model independent way demonstrate that the knee is caused by the 
decreasing flux of light primaries~\cite{muon}.  \\

Investigations of anisotropies in the arrival directions of the 
cosmic rays give additional information on the cosmic ray origin 
and of their propagation.  
Depending on the model of the origin of the knee and on 
the assumed structure of the galactic magnetic field
one expects large-scale anisotropies on a scale of 
$10^{-4}$ to $10^{-2}$ in the energy region of the knee.
The limits of large-scale anisotropy analyzing the KASCADE data 
are determined to be between $10^{-3}$ at 0.7 PeV primary energy 
and $10^{-2}$ at 6 PeV~\cite{Gmaier1}. 
These limits were obtained by investigations of 
the Rayleigh amplitudes and phases of the first harmonics. 
Taking into account possible nearby sources of galactic cosmic 
rays like the Vela Supernova 
remnant the limits of KASCADE already exclude 
particular model predictions.

The interest for looking to point sources in the KASCADE data sample 
arises from the possibility of unknown near-by sources, where the 
deflection of the charged cosmic rays would be small or by sources 
emitting neutral particles like high-energy gammas or neutrons. 
The scenario for neutrons is very interesting for KASCADE-Grande,
since the neutron decay length at these energies is in the order of
the distance to the Galactic center. 
In KASCADE case no significant excess was found~\cite{Gmaier2}. \\

\section{KASCADE-Grande: Cosmic rays goes metagalactic}

\vspace*{0.6cm}
{\centering
\begin{minipage}{120mm}
`...the observed spectrum is a superposition of the spectra of
particles of galactic and metagalactic origin.' \\
{\it from: On the size spectrum of extensive air showers (1959), G.V.
Kulikov and G.B. Khristiansen~\cite{knee}.}
\end{minipage}}
\vspace*{0.6cm}

The highest energies above the so called ankle at a few EeV
are believed to be exclusively of extragalactic origin. 
Thus, in the experimentally scarcely explored region between 
the first (proton) knee and the ankle there are two more 
peculiarities of the cosmic ray spectrum expected: 
(i) A knee of the heavy component which is 
either expected (depending on the model)
at the position of the first knee scaled with Z (the charge) 
or alternatively with A (the mass) of iron. 
(ii) A transition region from galactic to extragalactic origin 
of cosmic rays, where there is no theoretical reason for a 
smooth crossover in slope and flux. Dependent on the considered 
astrophysical model the second knee is allocated to case (i) or
(ii), respectively~\cite{haungs-tev}.  \\

\vspace*{0.1cm}
\begin{figure}[ht]
\centering
\begin{minipage}{75mm}
\centering
\includegraphics[width=75mm]{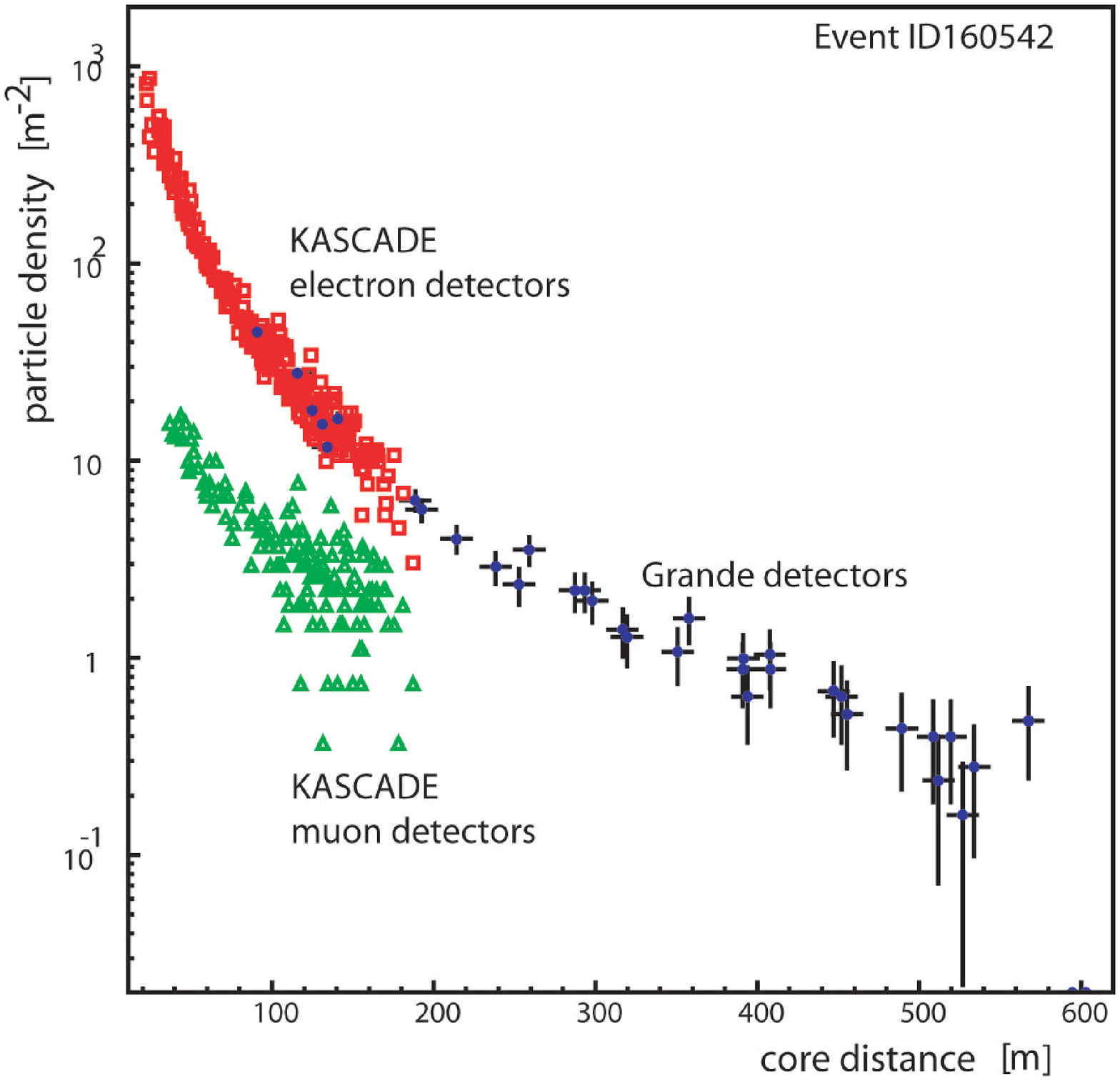}
\caption{\label{LDFexample}Particle densities in the different detector types 
of KASCADE-Grande measured for a single event.}
\end{minipage}\hspace{10mm}%
\begin{minipage}{75mm}
\centering
\includegraphics[width=75mm]{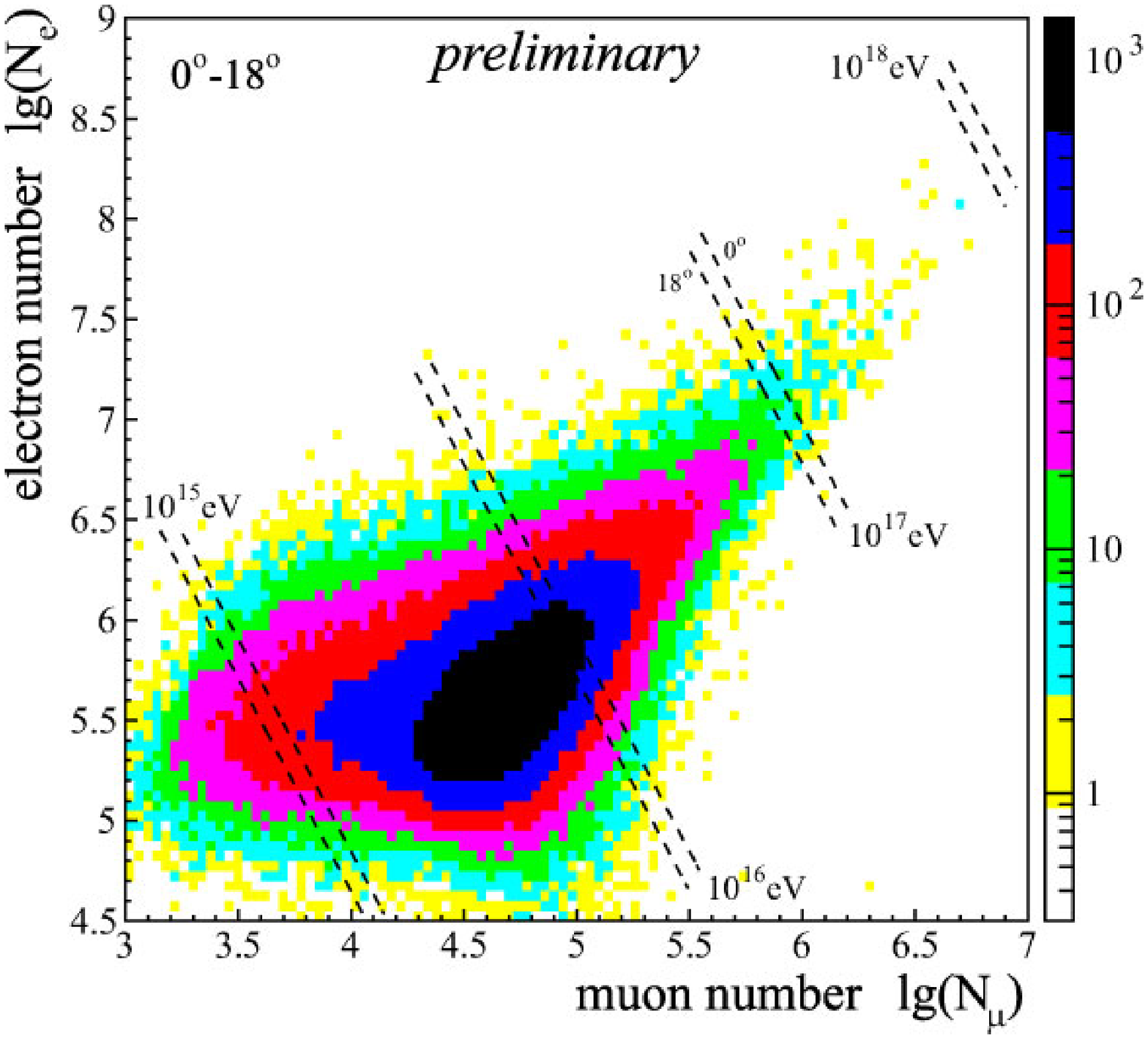}
\caption{\label{april04_data}Reconstructed electron vs. muon number
size spectrum after one year of data taking of
KASCADE-Grande~\cite{Glass05}.}
\end{minipage} 
\end{figure}
To distinguish between the two astrophysical scenarios shown in 
Figure~\ref{knee} constraints can be given by clarifying the 
existence and source of the second knee, which is possible 
by determining  the mass composition in the relevant energy 
range in detail. It is obvious that for the range 
between 50 and 1000 PeV sophisticated experiments are needed 
to measure the EAS with the same statistical and reconstruction 
accuracy as the KASCADE experiment discussed in the 
previous section. \\

Fig.~\ref{LDFexample} shows, for a single shower,
the lateral distribution of electrons and muons 
reconstructed with KASCADE and the charge particle densities 
measured by the Grande stations. This example illustrates the 
capabilities of KASCADE-Grande and the high quality of the data. 
The KASCADE-Grande reconstruction procedure follows iterative 
steps: shower core position, 
angle-of-incidence, and total number of charged particles are 
estimated from Grande array data~\cite{Glass05}; 
the muon densities and the reconstruction of the
total muon number are provided by the KASCADE muon 
detectors~\cite{vanBuren05}.
\begin{figure}[t]
\centering
\vspace*{0.1cm}
\includegraphics[width=100mm]{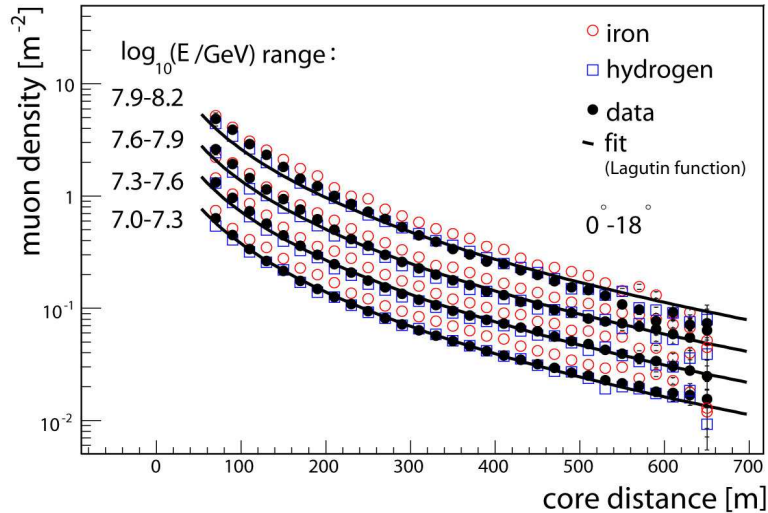}
\caption{Lateral density distributions of muons measured with 
KASCADE-Grande compared with simulated 
distributions~\cite{haungs-muon06}.} 
\label{muonldfs}
\end{figure}

In particular the possibility to reconstruct the total muon 
number for Grande measured showers is the salient feature of 
KASCADE-Grande compared to other experiments in this energy range.
To describe the lateral distribution (LDF) of the muons a 
Lagutin-like function is used.
In order to obtain stable fit results at single air shower
reconstruction, the shape parameter of the 
Lagutin-LDF is kept constant and only  
the muon number is estimated.

Figure~\ref{muonldfs} presents measured and simulated LDFs 
for different primary energies, where the energy has been 
roughly estimated by a linear combination of reconstructed total 
electron and muon numbers~\cite{haungs-muon06}.
Generally, the agreement between data and MC is very good 
and the LDF describes
the data reasonable well. \\

At the KASCADE experiment, the two-dimensional distribution 
shower size - truncated number of muons played the fundamental 
role in reconstruction of energy spectra of single mass groups. 
Hence, Figure~\ref{april04_data} illustrates the capability of
KASCADE-Grande to perform an unfolding procedure like in 
KASCADE. 
A hundred percent efficiency for KASCADE-Grande is reached for all 
primary particle types at energies above $2 \cdot 10^{16}\,$eV, 
thus providing a large overlap with the KASCADE energy range.

KASCADE-Grande started end of 2003 with combined measurements
of all detector components. 
Due to the fact that also for KASCADE-Grande 
a wealth of information on individual showers is available, 
tests of the hadronic interaction models and anisotropy studies 
will be possible in addition to the reconstruction of 
energy spectrum and composition.

\section{LOPES: Primed for the spectrum's end}

\vspace*{0.6cm}
{\centering
\begin{minipage}{120mm}
`Finally, we note that the mechanism of radio emission, and
especially the polarization, must be known to establish quantitative
correlations between the radio emission and other air shower
characteristics.' \\
{\it from: Detection of radio emission from extensive air showers
with a system of single half-wave dipoles (1967), S.N.
Vernov, G.B. Khristiansen et al.~\cite{vernov}.}
\end{minipage} }
\vspace*{0.6cm}

The traditional method to study extensive air showers (EAS) is to 
measure the secondary particles with sufficiently large particle 
detector arrays. In general these measurements provide only 
information on the actual status of the air shower cascade 
on the particular observation level. This hampers the determination 
of the properties of the EAS inducing primary as compared to 
methods like the observation of Cherenkov and fluorescence 
light~\cite{rpp}.
In order to reduce the statistical and systematic uncertainties of 
the detection and reconstruction of EAS, especially with respect 
to the detection of cosmic particles of highest energies, there is 
a current methodical discussion on new detection techniques. 
Due to technical restrictions in past times 
the radio emission accompanying cosmic ray air 
showers was a somewhat neglected EAS feature. For a review on the
early investigations of the radio emission in EAS in the 
60ties see~\cite{Allan71}. 
However, the study of this EAS component has experienced a 
revival by recent activities, in particular by the LOPES project. \\
\vspace*{0.1cm}
\begin{figure}[ht]
\centering
\begin{minipage}{75mm}
\centering
\includegraphics[width=75mm]{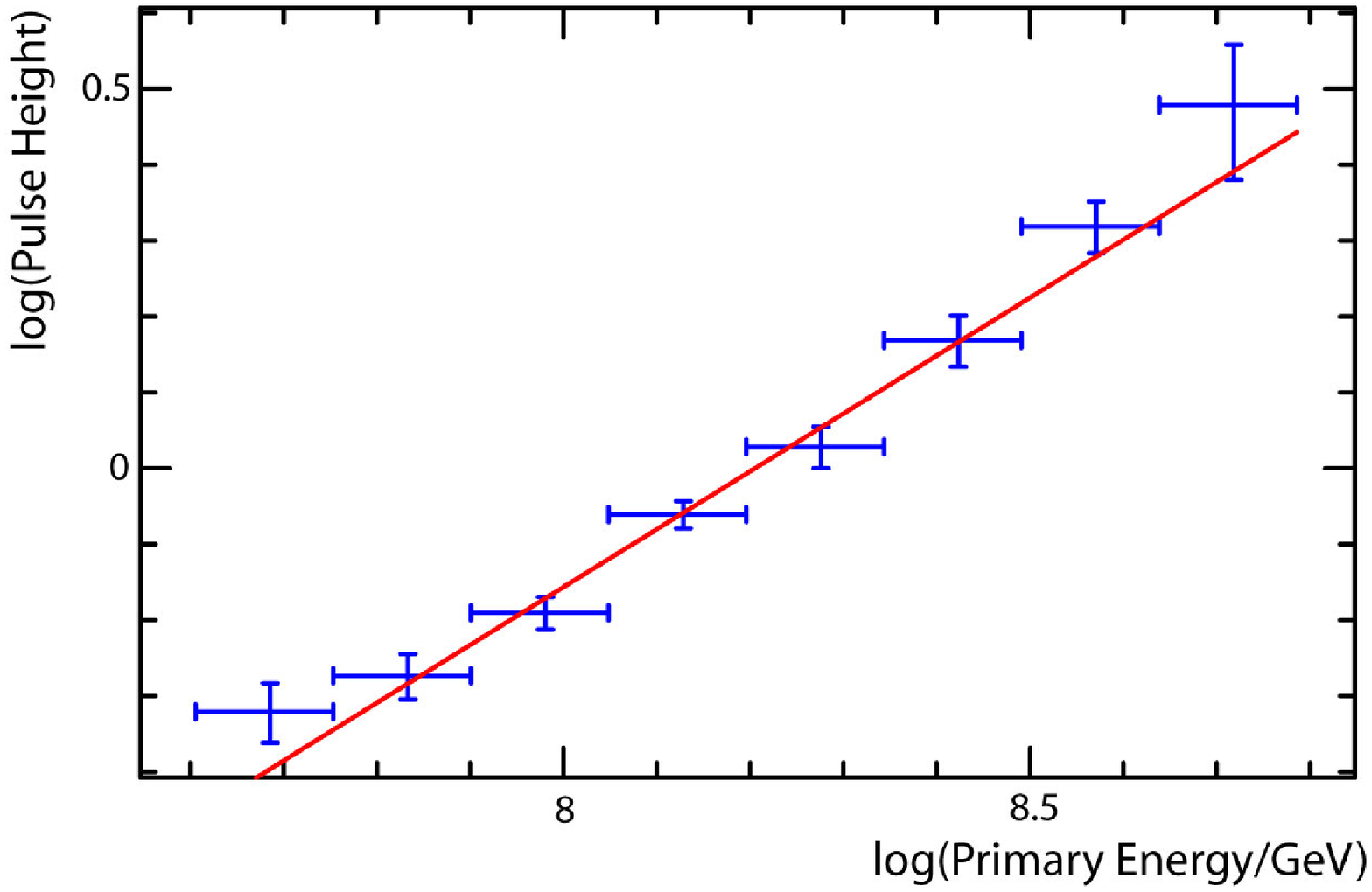}
\caption{Average radio pulse height of the LOPES-10 detected 
events (with shower core inside the KASCADE array and corrected for
the geomagnetic angle) plotted versus the 
primary particle energy as reconstructed by KASCADE~\cite{horneffer05}.}
\label{energy-10}
\end{minipage}\hspace{9mm}%
\begin{minipage}{75mm}
\centering
\includegraphics[width=75mm]{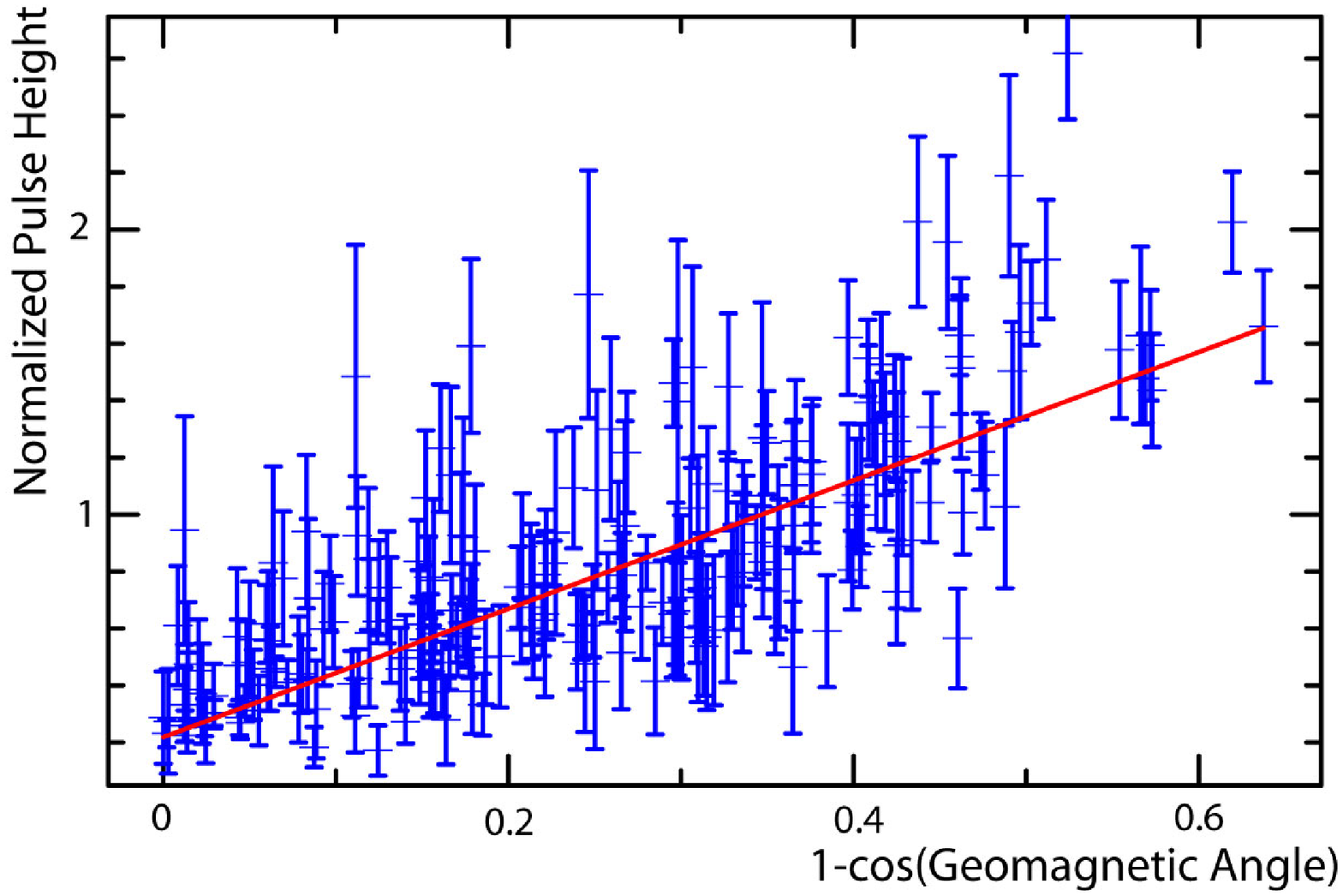}
\caption{Radio pulse height (corrected for muon number and 
distance to the shower axis) for individual air showers 
versus the cosine
of the angle to the geomagnetic field. The error bars are the
statistical errors~\cite{horneffer05}.}
\label{geo-10}
\end{minipage} 
\end{figure}

The main goal of the investigations in Karlsruhe in the frame of  
LOPES is the `calibration' of the shower radio emission 
in the primary energy range of $10^{16}\,$eV to $10^{18}\,$eV. 
I.e., to investigate in detail the correlation of the measured 
field strength with the shower parameters, in particular the 
orientation of the shower axis (geomagnetic angle, azimuth angle, 
zenith angle), the position of the observer (lateral extension 
and polarization of the radio signal), and the energy and mass 
(electron and muon number) of the primary particle.
In the following some first results of LOPES will be shown, 
obtained by a data set with 10 antennas (LOPES-10) installed. \\

In the frame of LOPES, also theoretical and detailed Monte Carlo 
studies of the radio emission 
are performed in the scheme of the so-called 
coherent geosynchrotron radiation.
Here, electron-positron pairs generated in the shower development 
gyrate in the Earth's magnetic field and emit radio pulses by
synchrotron emission~\cite{huege03,huege05}.
During the shower development the electrons and positrons  
are concentrated in a thin shower disk ($<2\,$m), 
which is smaller than one wavelength (at $100\,$MHz) of the 
emitted radio wave.
This situation provides the coherent emission of the radio signal. \\

When KASCADE-Grande has triggered LOPES the processing of the registered LOPES 
data includes several steps~\cite{horneffer05}. 
First, the relative instrumental delays are corrected using a 
known TV transmitter visible in the data. Next, the digital 
filtering, gain corrections and corrections of the trigger delays 
based on the known shower direction (from KASCADE) are applied and
noisy antennas are flagged. 
Then a time shift of the data 
and a correction for the azimuth and zenith
dependence of the antenna gain  
is done and the combination of the data is performed calculating
the resulting beam from all antennas. 
This digital beam forming allows to place a narrow antenna beam 
in the direction of the cosmic ray event.
To form the beam the data from each 
pair of antennas is multiplied time-bin by time-bin, the resulting 
values are averaged, and then the square root is taken while 
preserving the sign.
The resulting pulse is called the cross-correlation beam or CC-beam.
The finally obtained value $\epsilon_\nu$, which is the 
measured amplitude divided by the effective bandwidth, is
compared with further 
shower observables from KASCADE-Grande. \\

The LOPES-10 data set is subject of various analyses using different 
selections: With an event sample obtained by stringent cuts the proof of principle 
to detect air showers in the radio frequency range was 
given~\cite{Falck05}. 

\begin{figure}[ht]
\begin{minipage}{75mm}
\centering
\includegraphics[width=75mm]{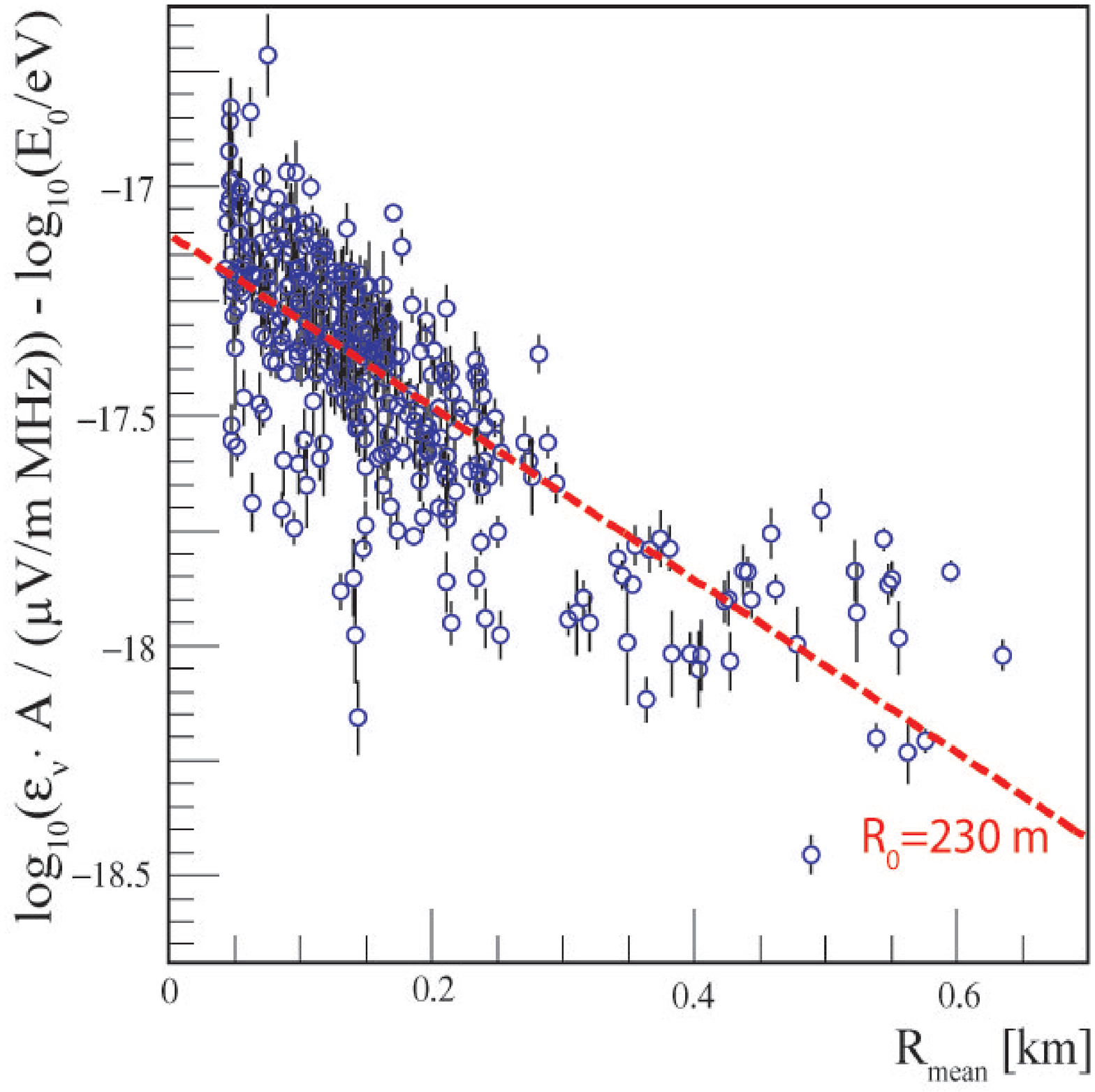}
\caption{Correlation (sample of distant events) 
of the pulse height corrected for primary energy 
with the mean distance of the shower axis to the radio antenna 
system. The line shows the result of a fit with an
exponential function~\cite{badea06}.}
\label{dist-10}
\end{minipage}\hspace{9mm}%
\begin{minipage}{75mm}
\centering
\includegraphics[width=75mm]{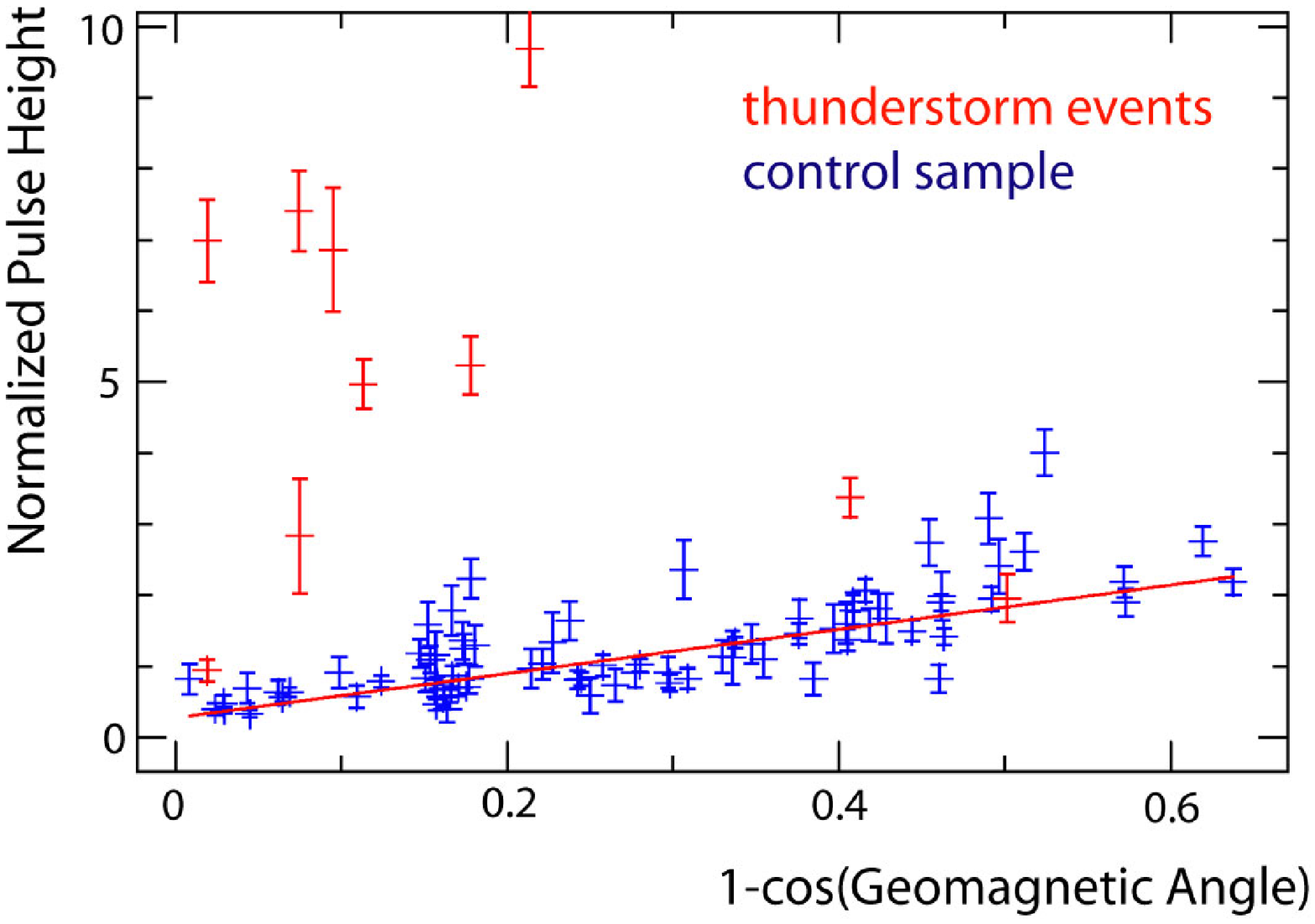}
\caption{Normalized pulse height of a control sample 
of detected events
and those detected during thunderstorms 
plotted against the geomagnetic angle.
The lines are fits to the data to describe the 
correlation~\cite{haungs-lopes06}.}
\label{thunder}
\end{minipage} 
\end{figure}
With showers fallen inside KASCADE the basic correlations with shower 
parameters are shown~\cite{lopes,horneffer05}.
As example, Fig~\ref{energy-10} depicts the dependence of the reconstructed 
averaged radio pulse height on the primary energy of the 
cosmic particles. 
The shown correlation supports 
the expectation that the field strength 
increases by a power-law with an index close to one
with the primary energy, i.e. that 
the received power of the radio signal increases quadratically 
with the primary energy of the cosmic rays. 

Fig.~\ref{geo-10} shows the correlation between the 
normalized reconstructed 
pulse height of the events with the geomagnetic angle. 
Normalized here means, that the detected pulse height is corrected 
for the dependence on the muon number, i.e.~to a large extent, 
the primary energy, and distance to the shower axis.
The clear correlation found suggests a geomagnetic origin for the
emission mechanism~\cite{lopes}. 

Besides the analyses of events with the core inside the antenna set-up,
KASCADE-Grande gives the possibility to search for distant events.
For each (large) shower triggering KASCADE, the information from 
the extension of KASCADE, i.e.~from the Grande array, is 
available.
From that information the shower can be reconstructed even if 
the core is outside the original KASCADE area, and a radio
signal can be searched for events which have distances up to 
$800\,$m from the center of the antenna set-up.
LOPES-10 detects clear EAS radio events at more than
$500\,$m distance from the shower axis for primary energies 
below $10^{18}\,$eV. That itself is a remarkable result, but
in addition, an important issue is the functional form of the 
dependence of the radio field strength with distance to the 
shower axis. 
After linear scaling of the pulse amplitude 
with the primary energy estimated by KASCADE-Grande a 
clear correlation with the mean distance of the shower axis to the
antennas is found (Fig.~\ref{dist-10}). 
This correlation can be described by an 
exponential function with a scaling radius in the order of a 
few hundred meters~\cite{badea06}.

Further interesting features 
are investigated with a sample of very inclined 
showers~\cite{petrovic06}. 
The sample is of special interest for a large scale application 
of this detection technique, as due to the low attenuation in the 
atmosphere also very inclined showers should be detectable with high 
efficiency. With LOPES one could show that events above $70^\circ$ 
zenith angle still emit a detectable radio signal.

Measurements during thunderstorms are of interest to 
investigate the role of the 
atmospheric electric field in the emission process. 
The contribution of
an electric field to the emission mechanism is examined both,
theoretically and experimentally. 
Two mechanisms of amplification of radio emission are considered: 
the acceleration radiation of the shower particles and the 
radiation from the current that is produced by ionization electrons 
moving in the electric field.
The LOPES data is sampled in events recorded during thunderstorms, 
periods of heavy cloudiness and periods of cloudless weather. 
The finding is that during thunderstorms the radio 
emission can be strongly enhanced (Fig.~\ref{thunder}). 
No amplified pulses were found 
during periods of cloudless sky or heavy cloudiness, 
suggesting that the electric field effect for radio air shower 
measurements can be safely ignored
during non-thunderstorm conditions~\cite{buitink06}. \\

Meanwhile in LOPES 30 LOFAR-type antennas are in operation
(and in addition 11 dual polarized STAR-antennas~\cite{gemmeke}). 
LOPES-30~\cite{nehls05} (see Fig.\ref{grande}) has now a maximum baseline of 
approximately 260\,m by the addition of 20 new antennas. 
Each single antenna is absolute calibrated using a commercial
reference antenna.
The array provides now a larger sampling area to
the radio signal of a single event compared with the 
original LOPES-10 set-up. 
This provides the possibility for a more detailed investigation 
of the radio signal on a single air shower basis, 
in particular of its lateral extension.   
In addition, during the LOPES-30 measurements, 
emphasis is put on monitoring environmental conditions by 
measuring the static electric field and by recording parameters 
of nearby weather stations. 
Atmospheric conditions, in particular E-field
variations during thunderstorms, probably influence the radio 
emission during the shower development, 
and the measurement of the radio pulses. 
By monitoring the environmental conditions, and comparing
them with the antenna noise level as well as with the detected air
shower radio signals, correlations will be investigated and 
corrected for.  Hence, finally the obtained pulse amplitude
can be directly compared with theoretical expectations for the 
radio field strength~\cite{huege05}.

\section{Conclusions}

\vspace*{0.5cm}
The KASCADE measurements demonstrate that the knee in the primary 
cosmic ray spectrum, positioned at few times $10^{15}\,$eV 
is originating from the depletion of the light elements, 
and that we should expect consequently further kinks of the 
spectrum  when the heavier elements are disappearing. 
In any case if this feature is Z- or A-dependent remains a 
question of great interest. 

The measurements give also evidence that cosmic rays of energies 
around the knee arrive our Earth isotropically. 
In addition the careful analyses of the KASCADE data on basis of 
current hadronic interaction models, unavoidably invoked for 
the interpretation, reveal the deficiencies of the present models. 
None of them is able to describe the data satisfactorily. 
The main uncertainties of the KASCADE results arise from such 
kind a `model' dependence. It is is also a consequence of  
the lack of accelerator data at relevant energies, especially 
in the forward direction, which could help to tune the model 
descriptions adequately. 

Actually, in spite of the success of recent sophisticated 
experiments like KASCADE, they provide only weak constraints 
for detailed astrophysical models of origin, acceleration 
and propagation of cosmic rays for explaining the discontinuities 
in the primary cosmic ray spectrum.  \\

The extension of KASCADE to KASCADE-Grande aims at the 
question if there do exist at higher energies knee-like 
structure associated with the disappearance of the heavier 
elements. This feature is expected to constrain more details 
of the astrophysical models and conjectures. 
KASCADE-Grande proceeds with the multi detector concept of 
the measurements in order also to investigate, 
in particular different aspects of the hadronic 
interaction models up to primary energies of $10^{18}\,$eV. \\

With setting up a small array of simple dipole antennas 
within the area of KASCADE-Grande, using sophisticated 
electronic and reconstruction procedures, it is attempted 
to develop a concept of air shower detection by the radio 
frequency emission during the shower development. 
First results obtained by correlating of the observed 
radio field strength with the shower parameters obtained 
by the KASCADE measurements appear to be very promising 
for a more detailed understanding of the emission mechanism 
from atmospheric showers.  
The main goal of the LOPES project with installing an 
increased number of antennas is the investigation of 
the relation between the radio emission from extensive 
air showers with the properties of the primary particles. 
Such studies are hoped to pave the way for a large-scale 
application of a novel technique in future cosmic ray 
experiments. \\

\vspace*{0.5cm}
{\noindent
{\bf Acknowledgments:} \\
\noindent
The author would like to acknowledge all members of the 
KASCADE, KASCADE-Grande, and LOPES collaborations for their 
deeply committed work on these sophisticated experiments.}

\smallskip

\end{document}